\documentclass[10pt,conference]{IEEEtran}
\usepackage{graphicx}    
\usepackage{amsfonts}    
\usepackage{booktabs}    
\usepackage{hyperref}    
\usepackage{subcaption}

\usepackage{float}
\usepackage{amsmath}
\usepackage{comment}

\begin{document}

\title{Entropy Estimation in Multi-Qutrit Systems via Variational and Classical Neural Networks}

\author{
\IEEEauthorblockN{Sai Sakunthala Guddanti\IEEEauthorrefmark{1}\IEEEauthorrefmark{2}\IEEEauthorrefmark{3},
Anil Prabhakar\IEEEauthorrefmark{2},
Ria Rushin Joseph\IEEEauthorrefmark{3}}

\IEEEauthorblockA{\IEEEauthorrefmark{1}Centre for Q. Info, Comm. and Computing, IIT Madras, Chennai 600036, India}

\IEEEauthorblockA{\IEEEauthorrefmark{2}Department of Electrical Engineering, IIT Madras, Chennai 600036, India}

\IEEEauthorblockA{\IEEEauthorrefmark{3}School of Information Technology, Deakin University, Burwood, Victoria 3125, Australia}
}

\maketitle 

\begin{abstract}
We present a systematic study of von Neumann entropy estimation in multi-qutrit quantum systems using two complementary approaches: variational quantum algorithms (VQAs) and classical convolutional neural networks (CNNs), evaluated using an ideal (noise-free) quantum simulator. For systems up to three qutrits, we construct and evaluate 11 hardware-efficient SU(3)-inspired ansatzes. A parameter sweep shows that estimation accuracy is primarily determined by the number of trainable parameters, provided sufficient entanglement is present. Based on this study, we fix the parameter count to approximately 120 for subsequent experiments, observing that increasing entangling-gate counts beyond a threshold yields only marginal improvements. For larger systems (two to five qutrits), we use a CNN trained on measurement outcomes from tensor-product mutually unbiased bases. The model achieves accurate and stable predictions and exhibits a systematic improvement in performance with system size, with the highest errors for two-qutrit systems and the lowest for five-qutrit systems. Notably, using only 12.5\% of the measurements required for full state tomography is sufficient to reach 90th-percentile absolute errors of approximately 0.13-0.16 nats for both four- and five-qutrit systems. The CNN model is also robust to shot noise and generalizes well to out-of-distribution states. Overall, within the simulated settings studied here, our results indicate a transition in practical methods: VQAs are effective for small systems, while CNN-based estimators offer improved scalability and robustness for larger qutrit systems
\end{abstract}

\begin{IEEEkeywords}
von Neumann entropy, qutrit systems, variational quantum algorithms, convolutional neural networks, mutually unbiased bases, ideal quantum simulation
\end{IEEEkeywords}

\section{Introduction}

Extracting useful information from an unknown quantum state is a central challenge in quantum information processing. While full quantum state tomography provides complete reconstruction of the density matrix, its experimental and computational cost scales exponentially with system size, rendering it impractical for even moderately large systems. In many applications, however, one is interested only in specific properties of the state, such as purity, fidelity, entanglement, or entropy. This has motivated the development of tomography-free methods for directly estimating physically relevant quantities.

Among these, the von Neumann entropy plays a fundamental role as a measure of mixedness, correlations, and information content. The von Neumann entropy of a quantum state $\rho$ is defined as
\[
S(\rho) = -\mathrm{Tr}(\rho \log \rho),
\]
which quantifies the mixedness of the state.
It is widely used in quantum thermodynamics, many-body physics, and the characterization of noisy intermediate-scale quantum (NISQ) devices. However, estimating entropy is inherently challenging due to its nonlinear dependence on the eigenvalue spectrum of the density matrix, making direct approaches difficult to scale.

While most prior work has focused on qubit systems, many experimental platforms—including superconducting circuits, trapped ions, and photonic systems—naturally support higher-dimensional quantum states. In particular, qutrits offer increased information density and can alter resource trade-offs in quantum algorithms. At the same time, their larger Hilbert space introduces additional challenges for state characterization, making scalable entropy estimation in qutrit systems an important and relatively unexplored problem.

In this work, we investigate tomography-free estimation of von Neumann entropy in multi-qutrit quantum systems using two complementary approaches. First, we study VQAs in low-dimensional regimes, where circuit structure and parameterization can be explicitly controlled. We design hardware-efficient SU(3)-inspired ansatzes built from Gell--Mann rotation (GMR) gates and qutrit SUM (CSUM) operations, and systematically analyze the interplay between parameter count, circuit depth, and entanglement in entropy estimation.

To address scalability beyond the reach of VQAs, we introduce a classical approach based on CNNs. This method operates directly on measurement statistics obtained from tensor-product mutually unbiased bases (MUBs), learning a mapping from measurement outcomes to entropy values without requiring density matrix reconstruction.

Together, these approaches allow us to explore complementary regimes: structured, interpretable quantum models in low dimensions, and scalable, data-driven classical models for larger systems. This combined framework enables a systematic investigation of performance, scaling behavior, and practical trade-offs in entropy estimation for multi-qutrit systems.

\section{Related work}

A wide range of methods have been proposed to estimate information-theoretic properties of quantum states without full tomography. These approaches address the challenges of estimating quantities such as the von Neumann entropy directly from limited access to quantum systems.

Existing work can be broadly categorized into four paradigms: (i) quantum query models, which assume structured or oracle-based access to the state; (ii) measurement-driven methods, which infer properties from sampled measurement data; (iii) machine-learning approaches, which learn mappings from data to target quantities; and (iv) variational quantum algorithms, which use parametrized circuits and classical optimization. In the following, we briefly review representative methods from each category.

A first class of entropy estimation approaches assumes structured access to the quantum state through oracle models, such as purified or controlled unitary access. Within this framework, entropy estimation is reduced to extracting spectral properties of the density matrix using quantum algorithms.

These methods typically evaluate nonlinear functions of the eigenvalues of $\rho$ via block-encoding, combined with polynomial transformations implemented using quantum signal processing or quantum singular value transformation (QSVT). The von Neumann entropy can then be obtained through trace estimation or related primitives.

Subramanyam \emph{et al.}~\cite{subramanyam} construct a block-encoding of $\rho^\alpha$ and estimate traces using DQC1-style protocols, while Wang \emph{et al.}~\cite{newentropyalgo} extend this framework to broader information-theoretic quantities without requiring spectral gap assumptions. Related polynomial-approximation methods have been studied for classical distributions~\cite{probabilitydistributions} and extend naturally to quantum settings. More recently, quantum phase processing (QPP)~\cite{phaseprocessing} reduces overhead by implementing polynomial transformations with a single ancilla qubit.

Despite strong theoretical guarantees, these approaches rely on idealized oracle access and deep circuits, limiting their applicability on near-term devices.

A second class of approaches estimates entropic quantities using independent copies of the quantum state and classical post-processing of measurement data. By avoiding coherent oracle access and full state reconstruction, these methods are well suited to experimental constraints on NISQ devices.

Early progress relied on randomized measurement protocols. Brydges \emph{et al.}~\cite{randommeasurements} showed that nonlinear quantities such as entropy can be estimated using random local unitaries and single-copy measurements, with values extracted from outcome correlations.

From a complexity perspective, Acharya \emph{et al.}~\cite{acharya} analyzed entropy estimation via Schur-basis measurements, establishing optimal copy complexity scaling. Related ideas underpin the classical shadow framework~\cite{shadows}, which enables efficient estimation of many observables from randomized measurements.
Polynomial-approximation methods using only state copies have also been proposed~\cite{quantumentropyalgo}, while the Samplizer framework~\cite{samplizer} connects sample-based and oracle-based approaches by implementing polynomial transformations without explicit block-encoding.

Overall, these methods replace idealized oracle access with experimentally feasible measurements, often at the cost of increased sample complexity or more involved measurements.

Machine learning approaches estimate quantum information measures directly from measurement data, bypassing explicit state reconstruction by learning nonlinear functionals of the quantum state.

Wang and Davis~\cite{neuralautoregressive} introduced neural autoregressive quantum states (NAQS), which learn compact probabilistic representations enabling estimation of information-theoretic quantities. Subsequent work focused on direct prediction from measurements: Koutný \emph{et al.}~\cite{deeplearningentangle} demonstrated that deep networks can estimate entanglement from incomplete data with robustness to measurement settings.

More recent supervised approaches map measurement outcomes to entropy with high sample efficiency and uncertainty quantification~\cite{supervisedlstm}, while generative models~\cite{renyignn} learn effective distributions to estimate such quantities. Scaling to larger systems, Wang \emph{et al.}~\cite{entanglementusingml} proposed methods for multipartite entanglement estimation using only local measurements.

Overall, these methods highlight the ability of machine learning to approximate nonlinear information-theoretic functionals from data, though most work focuses on qubits and entanglement, with limited exploration of entropy estimation in higher-dimensional systems such as qutrits.

Variational approaches recast entropy estimation as an optimization problem over auxiliary operators or states, avoiding explicit diagonalization~\cite{Berta2017}. Building on this, variational quantum algorithms (VQAs) use parameterized circuits and classical optimization to approximate functions of $\rho$, enabling entropy estimation from partial information~\cite{PhysRevE.109.044117}. More general formulations based on information-theoretic objectives, such as Donsker--Varadhan representations, allow multiple quantities to be estimated within a unified framework~\cite{PhysRevA.109.032431}.

Related methods include variational disentangling, which concentrates correlations into subsystems for efficient estimation~\cite{disentanglingqnn}, and extensions to quantities such as fidelity and trace distance~\cite{VQAforfidelity}. Purity-based formulations provide a unified route to multiple measures while mitigating optimization challenges such as barren plateaus~\cite{VQApurity}, and recent work extends these techniques to relative entropy using quadrature-based approximations compatible with near-term devices~\cite{estimatingquantumrelativeentropies}.

Overall, variational methods offer a flexible and hardware-compatible framework for estimating nonlinear information measures when direct spectral approaches are impractical.

\section{Methodology}

\subsection{Variational Quantum Algorithm (VQA) Approach}

We develop a variational framework for estimating the von Neumann entropy of multi-qutrit quantum states. While most prior variational approaches focus on qubit systems, extending these ideas to higher-dimensional settings introduces new challenges in circuit design and scalability.

The central idea is to learn a measurement basis in which the entropy can be inferred directly from classical measurement statistics. The variational circuit preserves the intrinsic entropy of the state and instead seeks a unitary transformation that approximately aligns the computational basis with the eigenbasis of the density matrix.

Let $\rho \in \mathbb{C}^{d^n \times d^n}$ denote an $n$-qutrit state with $d=3$. A parameterized unitary $U(\theta)$, implemented using GMR gates and CSUM operations, defines the transformed state
\begin{equation}
\rho(\theta) = U(\theta)\, \rho \, U^\dagger(\theta).
\end{equation}
Projective measurements in the computational basis $\{ |i\rangle \}$ yield a probability distribution
\begin{equation}
p_i(\theta) = \langle i | \rho(\theta) | i \rangle.
\end{equation}

The measured probability vector $p_i$
is the diagonal of $U\rho U^\dagger$ in computational basis. By the Schur--Horn theorem, this diagonal vector is majorized by the eigenvalue vector $\lambda(\rho)$. Since the Shannon entropy is Schur-concave, it follows that,
\[
H(p)\geq H(\lambda(\rho))=S(\rho),
\]
with equality when the measurement basis is the eigenbasis of $\rho$.

To learn an informative measurement basis, we maximize the Kullback--Leibler (KL) divergence between $p(\theta)$ and the uniform distribution $u_i = 1/d^n$, i.e.,
\begin{equation}
\mathcal{L}(\theta) = - D_{\mathrm{KL}}(p(\theta) \,\|\, u).
\end{equation}
Since
\[
D_{\mathrm{KL}}(p \,\|\, u) = \log(d^n) - H(p),
\]
this is equivalent to minimizing the Shannon entropy of the measurement distribution.

The optimization therefore favors bases that produce maximally non-uniform outcomes. For a fixed state, this occurs when the measurement basis aligns with the eigenbasis of $\rho$, in which case $p(\theta)$ approximates the eigenvalue spectrum. The resulting measurement entropy then provides an estimate of the von Neumann entropy of the quantum state.

\subsubsection{Qutrit Variational Circuit Design}

Circuit construction is implemented using the qudit library~\cite{seksaria2025qudit}. Single-qutrit operations are implemented using GMR gates, which generate SU(3) transformations analogous to Pauli rotations in qubit systems. Entanglement is introduced via the CSUM gate, a higher-dimensional analogue of the CNOT gate. Fixed Fourier-like (Hadamard-type) transformations are used as basis-change operations.

For a single qutrit, the variational ansatz consists of nine GMR gates, forming a complete basis of $\mathrm{SU}(3)$. This ensures full expressibility and enables entropy estimation with errors below $10^{-5}$ nats across a range of mixed states. Consequently, no further architectural variation is required in this setting.

We therefore focus on multi-qutrit systems, where circuit design must balance local expressibility, entanglement generation, and parameter efficiency. In particular, we study a family of three-qutrit ansatzes to systematically analyze these trade-offs.

\subsubsection{Ansatz Design and Parameter Efficiency}

A parameter study was conducted to quantify the effect of model capacity on entropy estimation accuracy. Using a fixed reference ansatz and a representative mixed quantum state, the number of trainable parameters was varied by increasing the number of circuit layers.

This controlled parameter sweep showed that, for a fixed circuit structure, increasing the number of trainable parameters leads to faster convergence and lower estimation error, indicating that model performance is strongly influenced by total parameter count, provided sufficient entanglement is present.

Motivated by this observation, we isolate architectural effects by fixing the total number of trainable parameters across all ansatzes. This is achieved by varying the number of circuit layers, enabling a controlled comparison between different circuit structures independent of parameter count.

Under this setting, we study ansatzes that vary in (i) the composition of GMR gates, (ii) gate ordering within each layer, and (iii) the placement of parameters relative to entangling operations. Each layer consists of local GMR rotations interleaved with CSUM gates, with fixed single-qutrit Fourier like gate operations (Hadamard) applied at the beginning and end of the layer.

\begin{figure*}[!t]
    \centering
    \begin{subfigure}{0.42\textwidth}
        \includegraphics[width=\linewidth, trim = 0.29cm 0.5cm 1cm 0.25cm, clip]{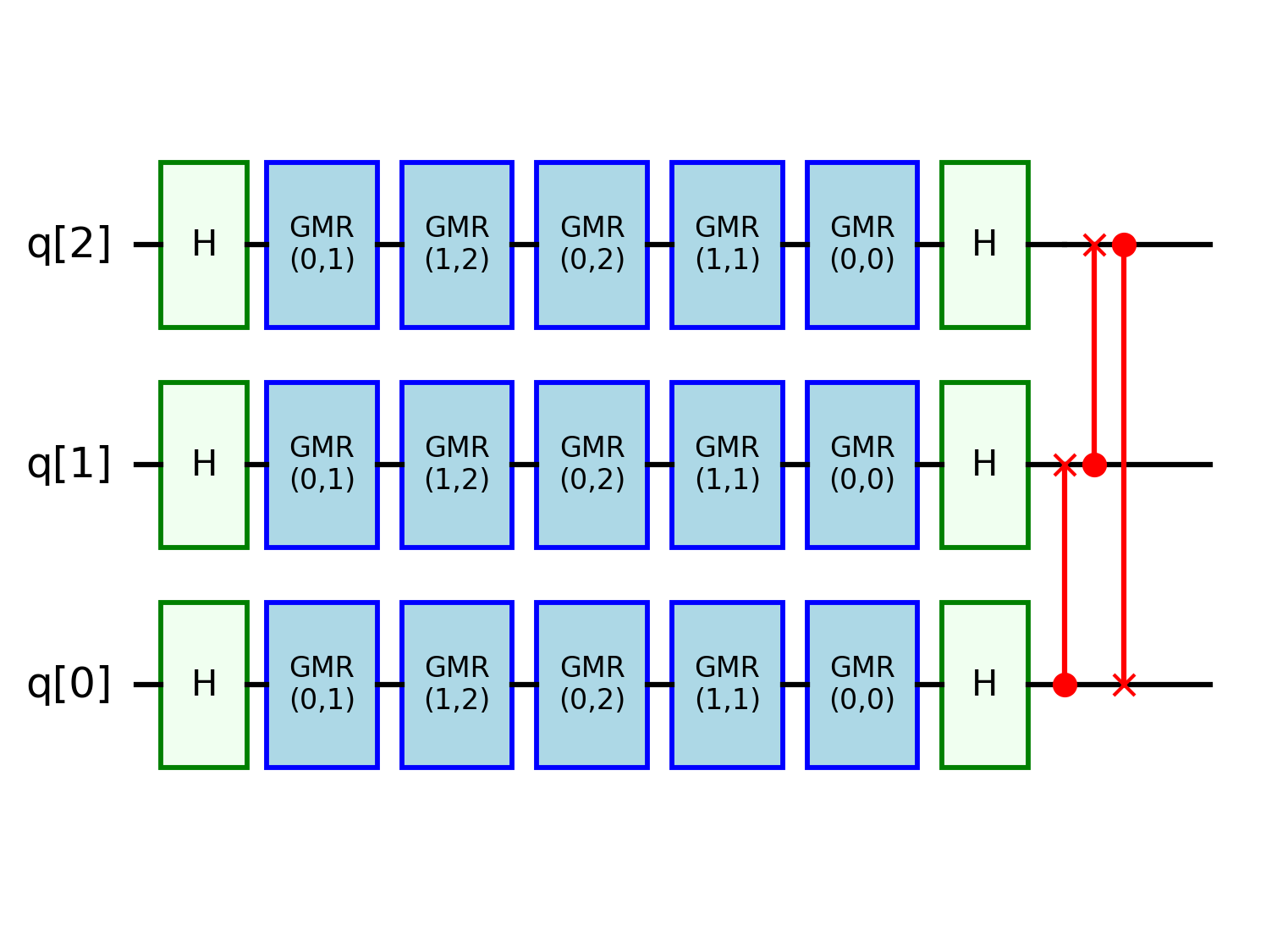}
        \caption{A1: $\left[(S^3 D^2) E\right]^8$}
        \label{fig:a1_ansatz}
    \end{subfigure}
    \hfill
    \begin{subfigure}{0.54\textwidth}
        \includegraphics[width=\linewidth, trim = 0.23cm 0.5cm 1.5cm 0.5cm, clip]{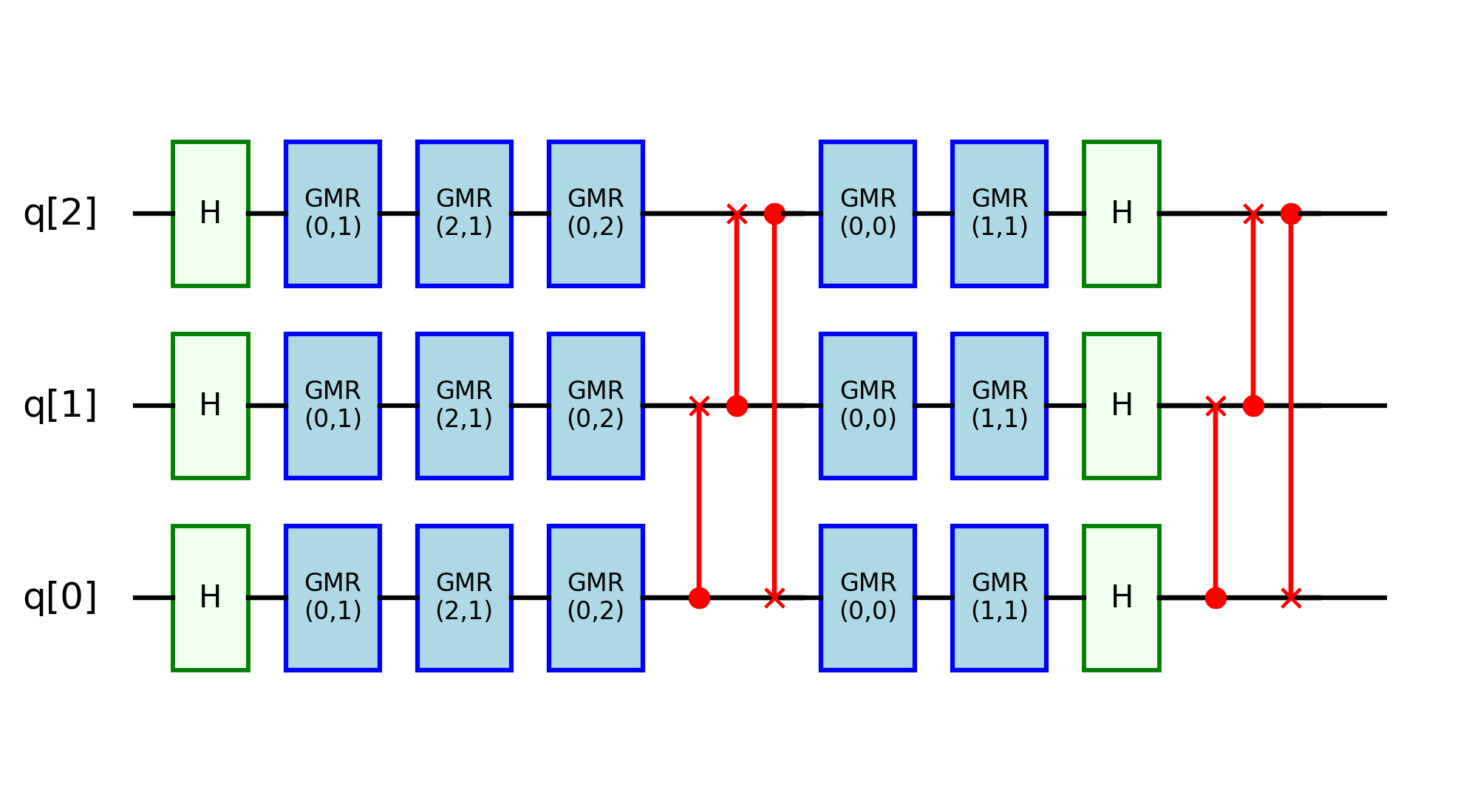}
        \caption{A4: $\left[(S^2 A^1 D^2) E\right]^8$}
        \label{fig:a4_ansatz}
    \end{subfigure}
    \begin{subfigure}{0.42\textwidth}
        \includegraphics[width=\linewidth, trim=0.25cm 0.5cm 1cm 0.25cm, clip]{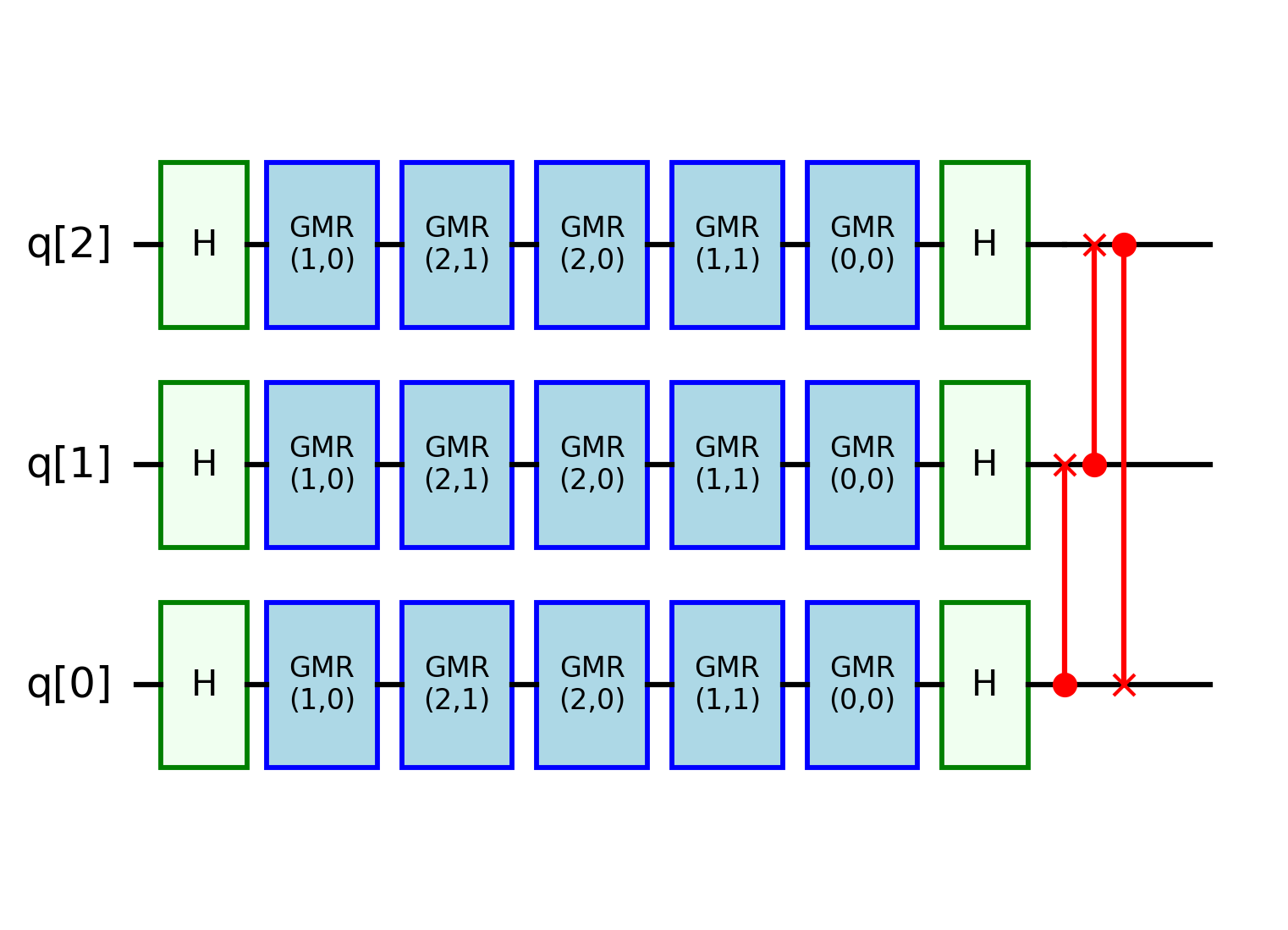}
        \caption{A5: $\left[(A^3 D^2) E\right]^8$}
        \label{fig:a5_ansatz}
    \end{subfigure}
    \hfill
    \begin{subfigure}{0.42\textwidth}
        \includegraphics[width=\linewidth, trim=0.25cm 0.5cm 1cm 0.25cm, clip]{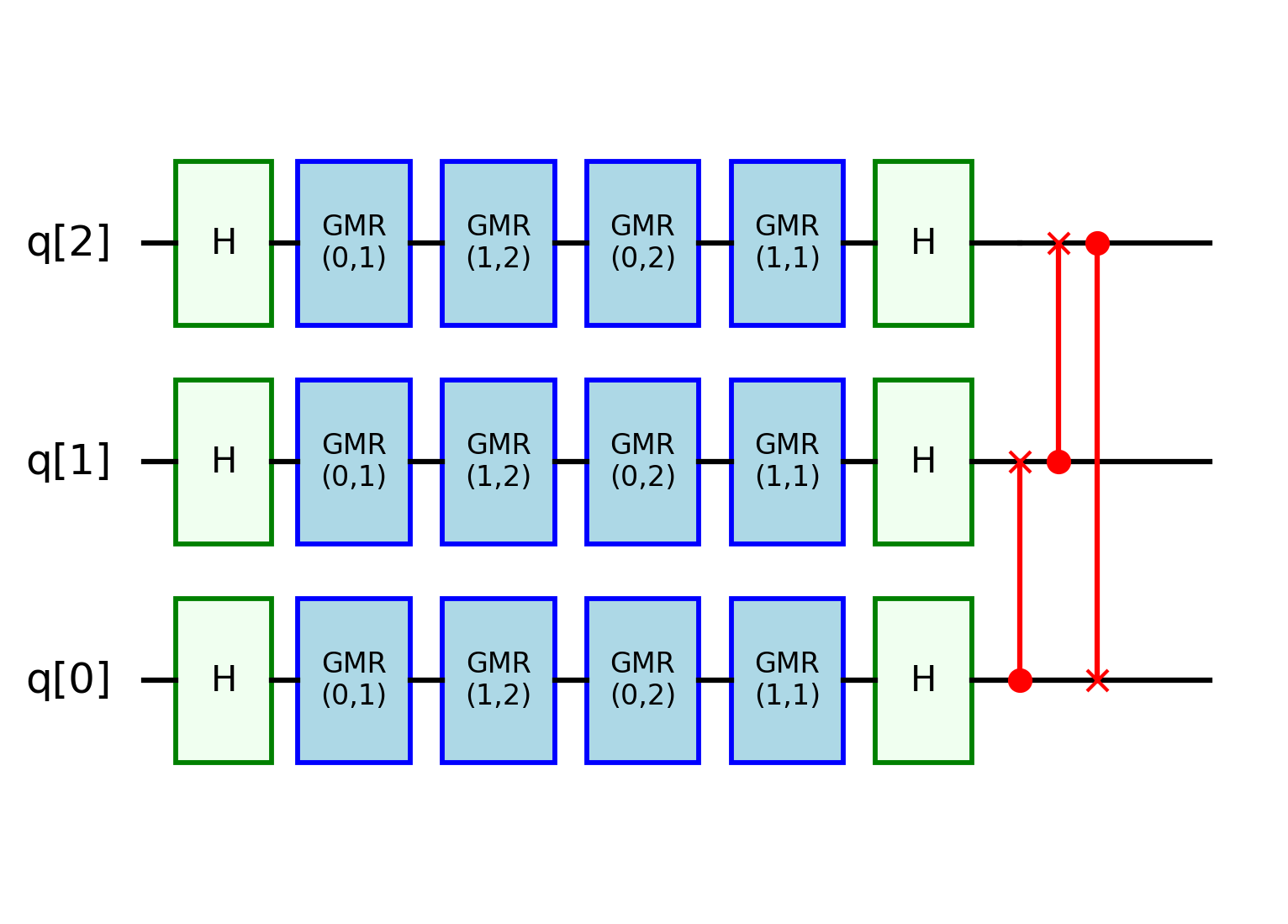}
        \caption{A6: $\left[(S^3 D) E\right]^{10}$}
        \label{fig:a6_ansatz}
    \end{subfigure}
    \begin{subfigure}{0.42\textwidth}
        \includegraphics[width=\linewidth, trim=0.25cm 0.5cm 1cm 0.25cm, clip]{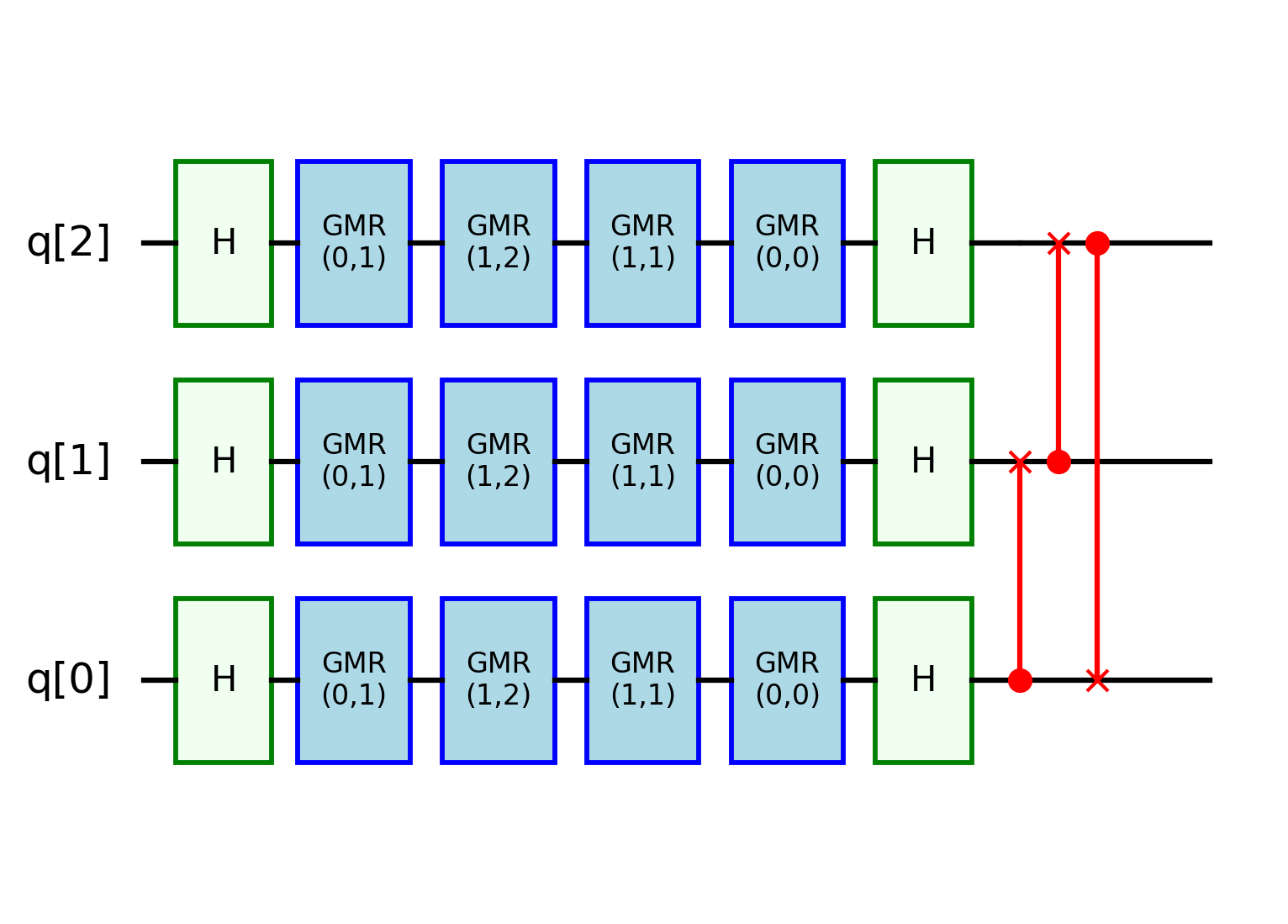}
        \caption{A7: $\left[(S^2 D^2) E\right]^{10}$}
        \label{fig:a7_ansatz}
    \end{subfigure}
    \hfill
    \begin{subfigure}{0.42\textwidth}
        \includegraphics[width=\linewidth, trim=0.25cm 0.5cm 1.5cm 0.25cm, clip]{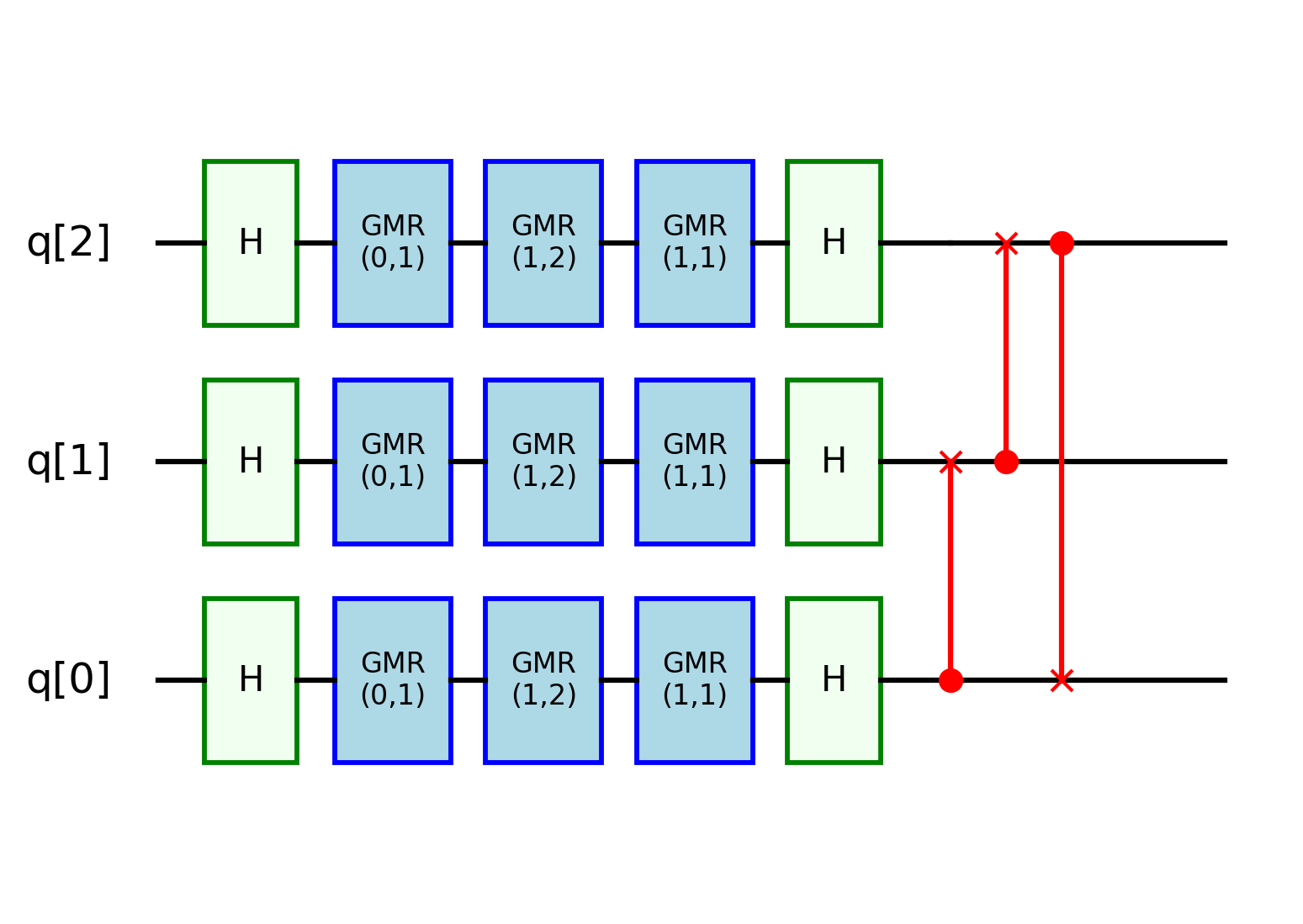}
        \caption{A8: $\left[(S^2 D) E\right]^{13}$}
        \label{fig:a8_ansatz}
    \end{subfigure}
    \vspace{0.5cm}
    \caption{Three-qutrit variational circuit ansatzes A1, A4, A5, A6, A7, and A8 are shown for single layer. Each circuit consists of layers of local GMR gates interleaved with entangling operations (CSUM gates). In the figures, the ordered index pairs $(i,j)$ label the generator class: $i<j$ denotes symmetric generators, $i>j$ antisymmetric generators, and $i=j$ diagonal generators; this notation is used solely to distinguish generator types. Each ansatz is denoted as $\left[\mathcal{B}\right]^L$, where $\mathcal{B}$ represents a variational block composed of symmetric ($S$), antisymmetric ($A$), and diagonal ($D$) GMRs followed by an entangling layer ($E$), and $L$ is the number of layers.}
    \vspace{1cm}
    \label{fig:3qutrit_ansatz_group1}
\end{figure*}

\begin{figure*}[!t]
    \centering
    \begin{subfigure}{0.59\textwidth}
        \includegraphics[width=\linewidth, trim=0.25cm 0.8cm 0.25cm 0.25cm, clip]{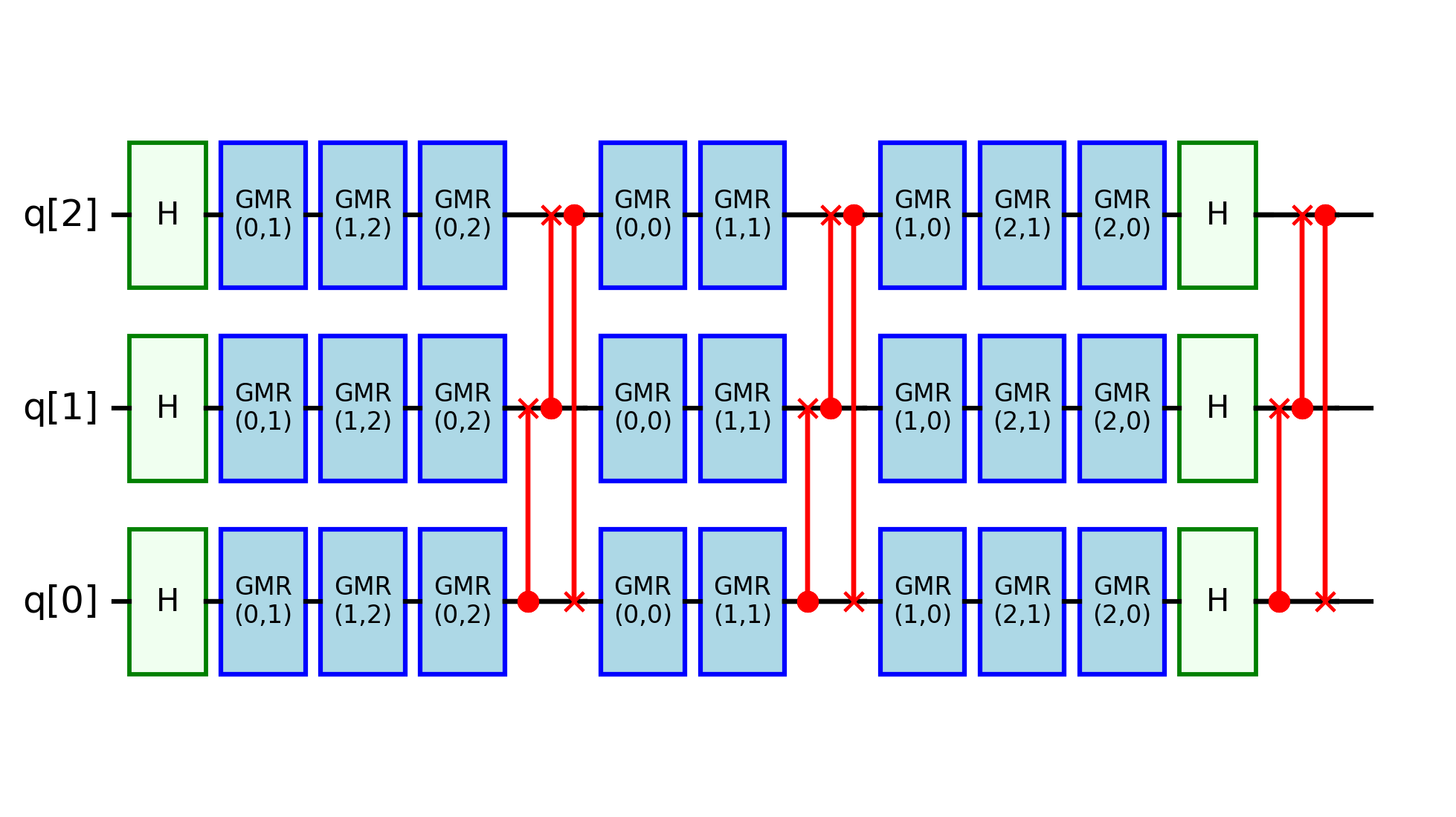}
        \vspace{-0.7cm}
        \caption{A2: $\left[(S^3 D^2 A^3) E\right]^5$}
        \label{fig:a2_ansatz}
    \end{subfigure}
    \hfill
    \begin{subfigure}{0.4\textwidth}
        \includegraphics[width=\linewidth, trim=1.5cm 0.8cm 0.5cm 0.25cm, clip]{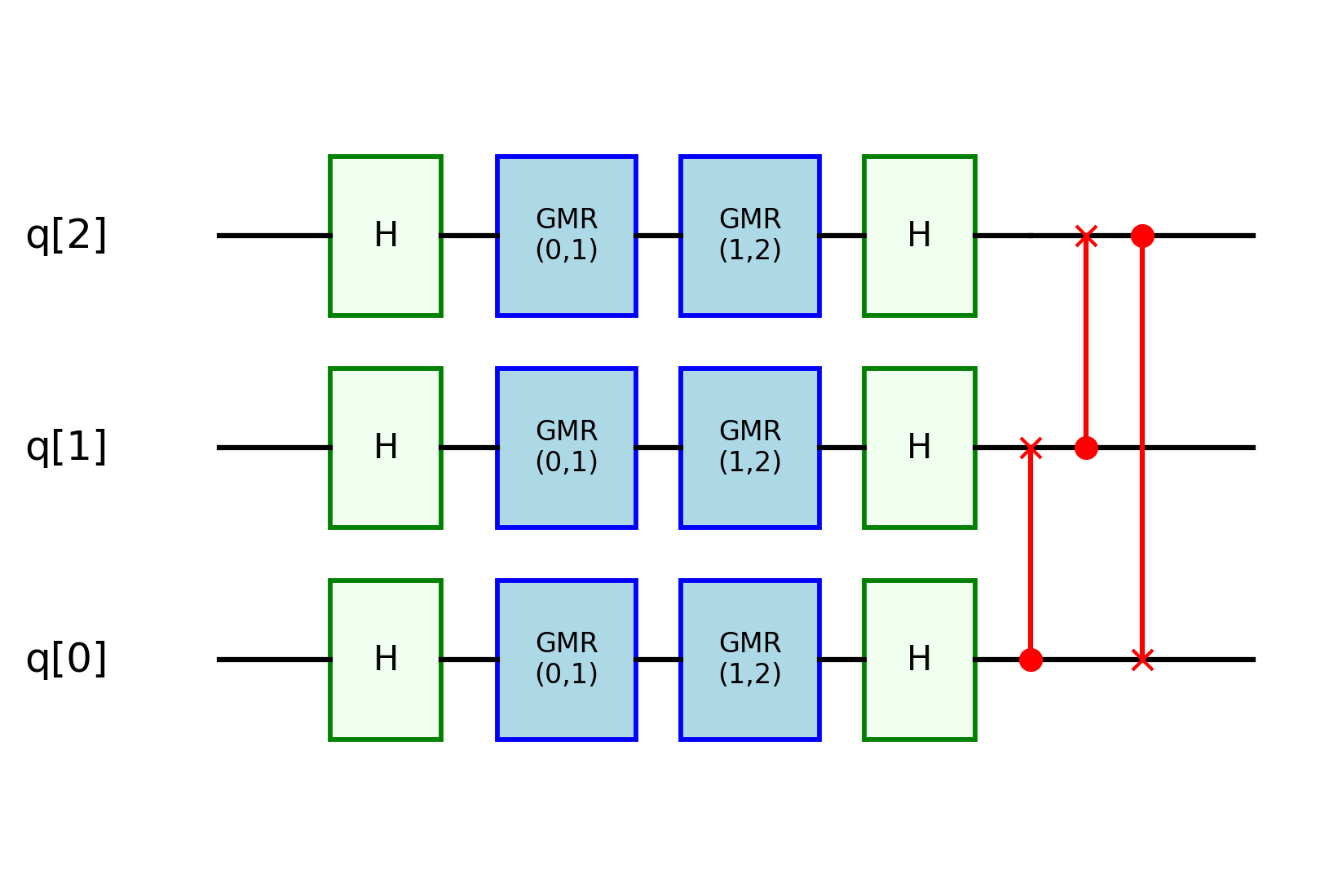}
        \vspace{-0.7cm}
        \caption{A11: $\left[(S^2) E\right]^{20}$}
        \label{fig:a11_ansatz}
    \end{subfigure}
    \hfill
    \begin{subfigure}{\linewidth}
        \includegraphics[width=\textwidth]{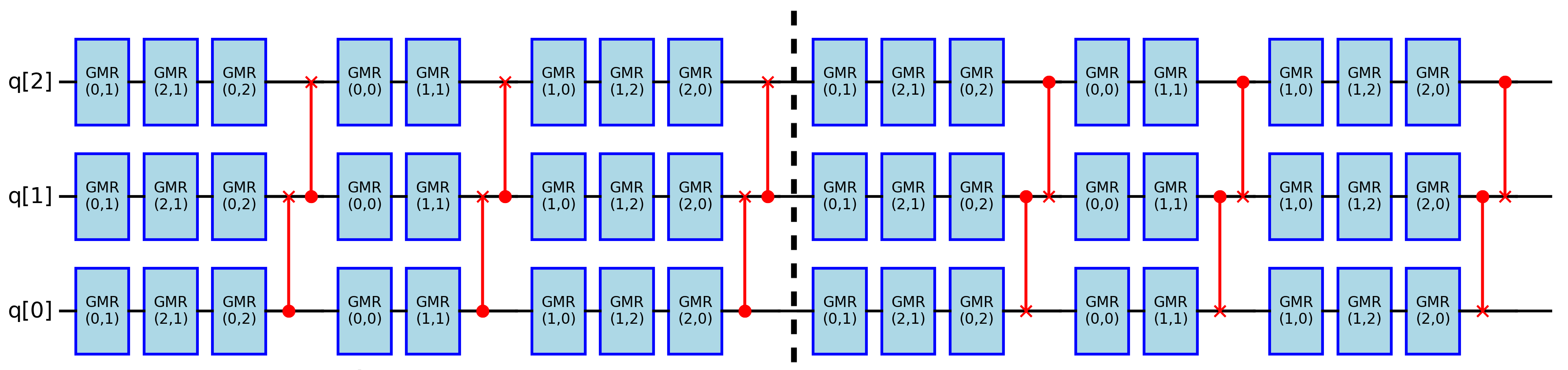}
        \caption{A3: $\left[(S^3 D^2 A^3) E_{\mathrm{alt}}\right]^5$}
        \label{fig:a3_ansatz}
    \end{subfigure}
    \hfill
    \begin{subfigure}{0.82\linewidth}
        \includegraphics[width=\textwidth, trim = 0cm 0.8cm 0cm 0.5cm, clip]{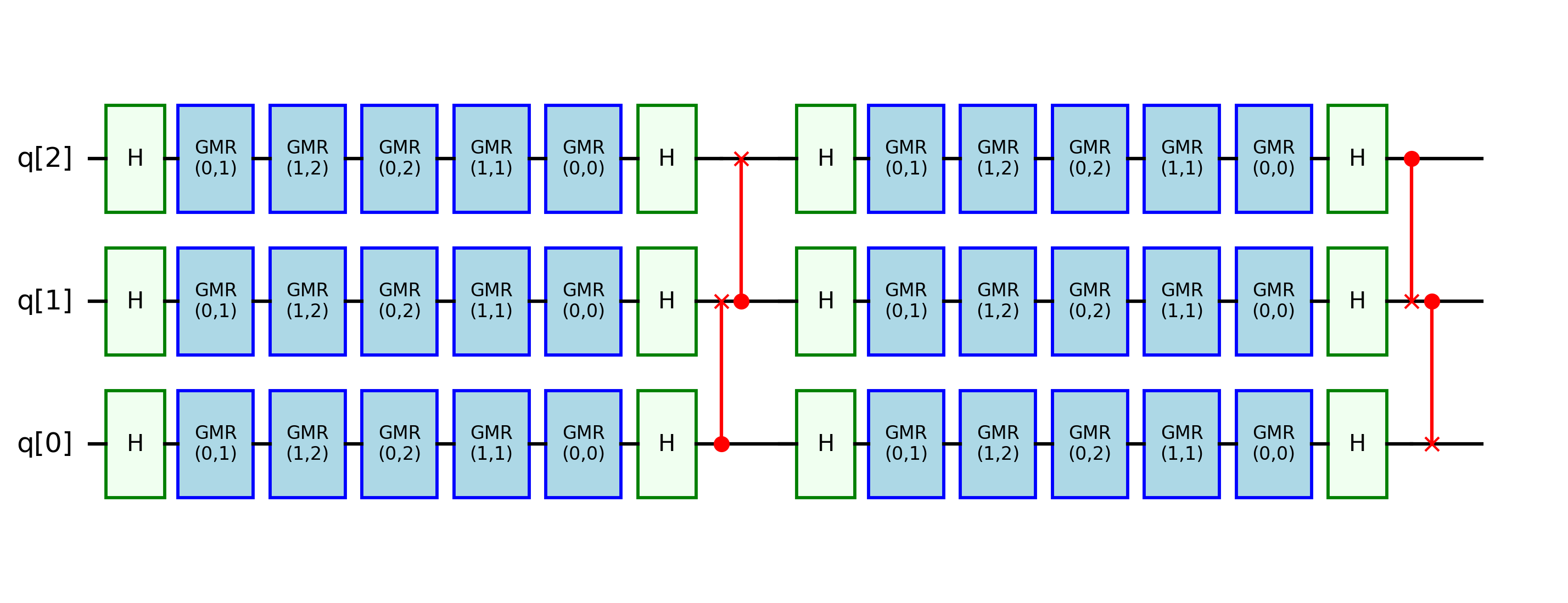}
        \caption{A9: $\left[(S^3 D^2) E_{\mathrm{alt}}\right]^8$}
        \label{fig:a9_ansatz}
    \end{subfigure}
    \hfill
    \begin{subfigure}{0.72\linewidth}
        \includegraphics[width=\textwidth, trim = 0.25cm 1cm 0cm 0.5cm, clip]{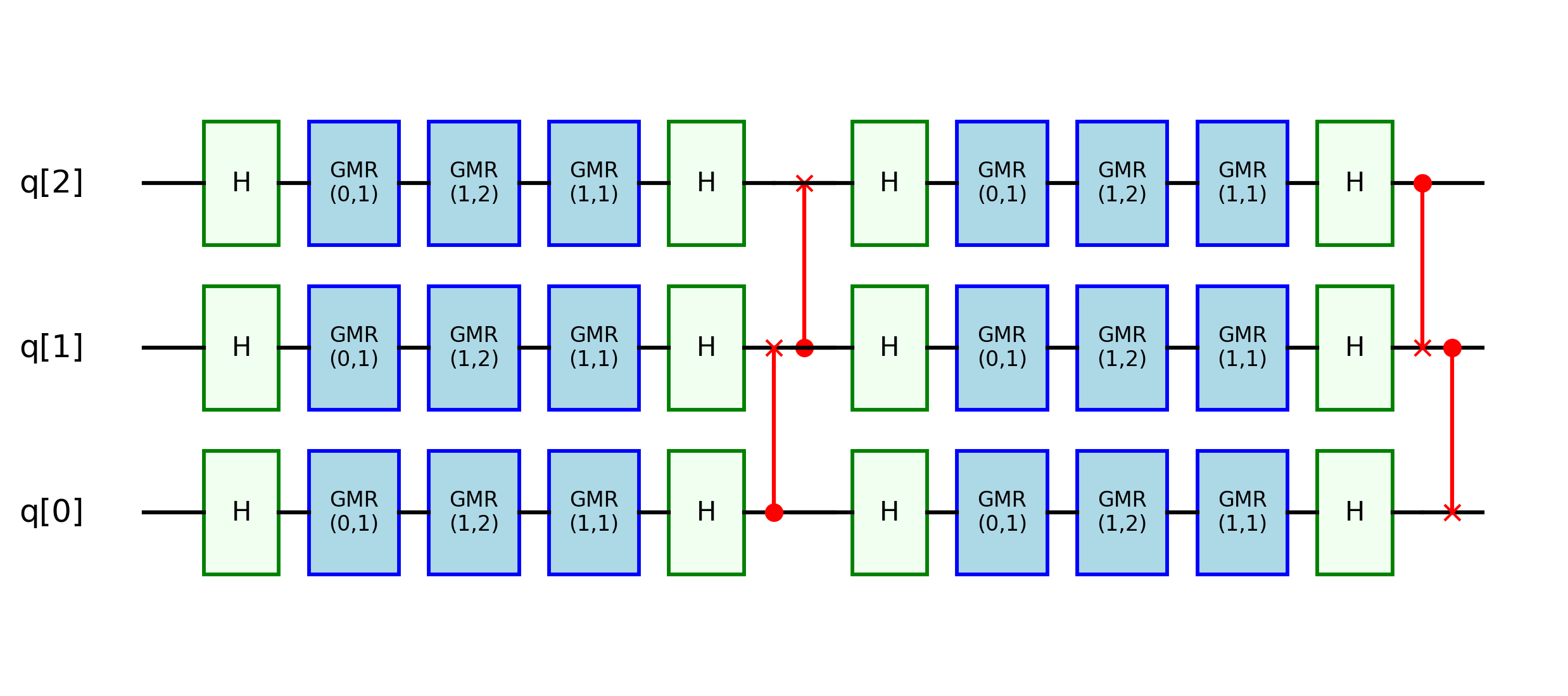}
        \caption{A10: $\left[(S^2 D) E_{\mathrm{alt}}\right]^{13}$}
        \label{fig:a10_ansatz}
    \end{subfigure}
    \caption{Three-qutrit variational circuit ansatzes A2, A11, A3, A9 and A10. Circuits follow the same GMR index notation as in Fig.~\ref{fig:3qutrit_ansatz_group1}. Ansatzes A3, A9 and A10 are shown over two layers to highlight alternating entanglement connectivity, while A2 and A11 are shown for a single layer. Hadamard-type basis-change gates are omitted for clarity in A3.
}
    \label{fig:3qutrit_ansatz_group4}
\end{figure*}

The following families of three-qutrit ansatzes were examined (see Figs.~\ref{fig:a1_ansatz}--\ref{fig:a11_ansatz}). Each ansatz is denoted as $\left[\mathcal{B}\right]^L$, where $\mathcal{B}$ represents a variational block composed of symmetric ($S$), antisymmetric ($A$), and diagonal ($D$) GMRs followed by an entangling layer ($E$), and $L$ is the number of layers :

\begin{itemize}
    \item \textbf{A1 (8 layers): $\left[(S^3 D^2) E\right]^8$} Baseline with five parameters per qutrit per layer (symmetric + diagonal GMRs) and full entangling layers.

    \item \textbf{A2, A3 (5 layers): $\left[(S^3 D^2 A^3) E\right]^5, \left[(S^3 D^2 A^3) E_{\mathrm{alt}}\right]^5$} Higher-expressibility designs with eight parameters per qutrit per layer. A3 additionally incorporates alternating entanglement connectivity across layers.

    \item \textbf{A4, A5 (8 layers): $\left[(S^2 A^1 D^2) E\right]^8, \left[(A^3 D^2) E\right]^8$} Variants of A1 with five parameters per qutrit per layer, differing in the choice of GMR generators and entangling structure.

    \item \textbf{A6, A7 (10 layers): $\left[(S^3 D) E\right]^{10}, \left[(S^2 D^2) E\right]^{10}$} Parameter-reduced variants (four parameters per qutrit per layer) derived from A1, with increased depth to compensate for reduced expressibility.

    \item \textbf{A8, A10 (13 layers): $\left[(S^2 D) E\right]^{13}, \left[(S^2 D) E_{\mathrm{alt}}\right]^{13}$} Strongly parameter-reduced designs (three parameters per qutrit per layer) with modified entanglement patterns and increased depth. A10 has reduced entangled gates because of alternating connectivity.

    \item \textbf{A9 (8 layers): $\left[(S^3 D^2) E_{\mathrm{alt}}\right]^8$} A1-like architecture with reduced entanglement per layer and alternating connectivity.

    \item \textbf{A11 (20 layers): $\left[(S^2) E\right]^{20}$} Minimal-parameter design (two parameters per qutrit per layer) requiring increased depth to match total parameter count.
\end{itemize}

All ansatzes A1--A7, A9, and A11 were constructed with 120 trainable parameters, while A8 and A10 contain 117. This controlled design enables systematic comparison of expressibility, parameter distribution, and entanglement structure.

\begin{figure*}[h]
    \centering
    \includegraphics[width=0.8\linewidth]{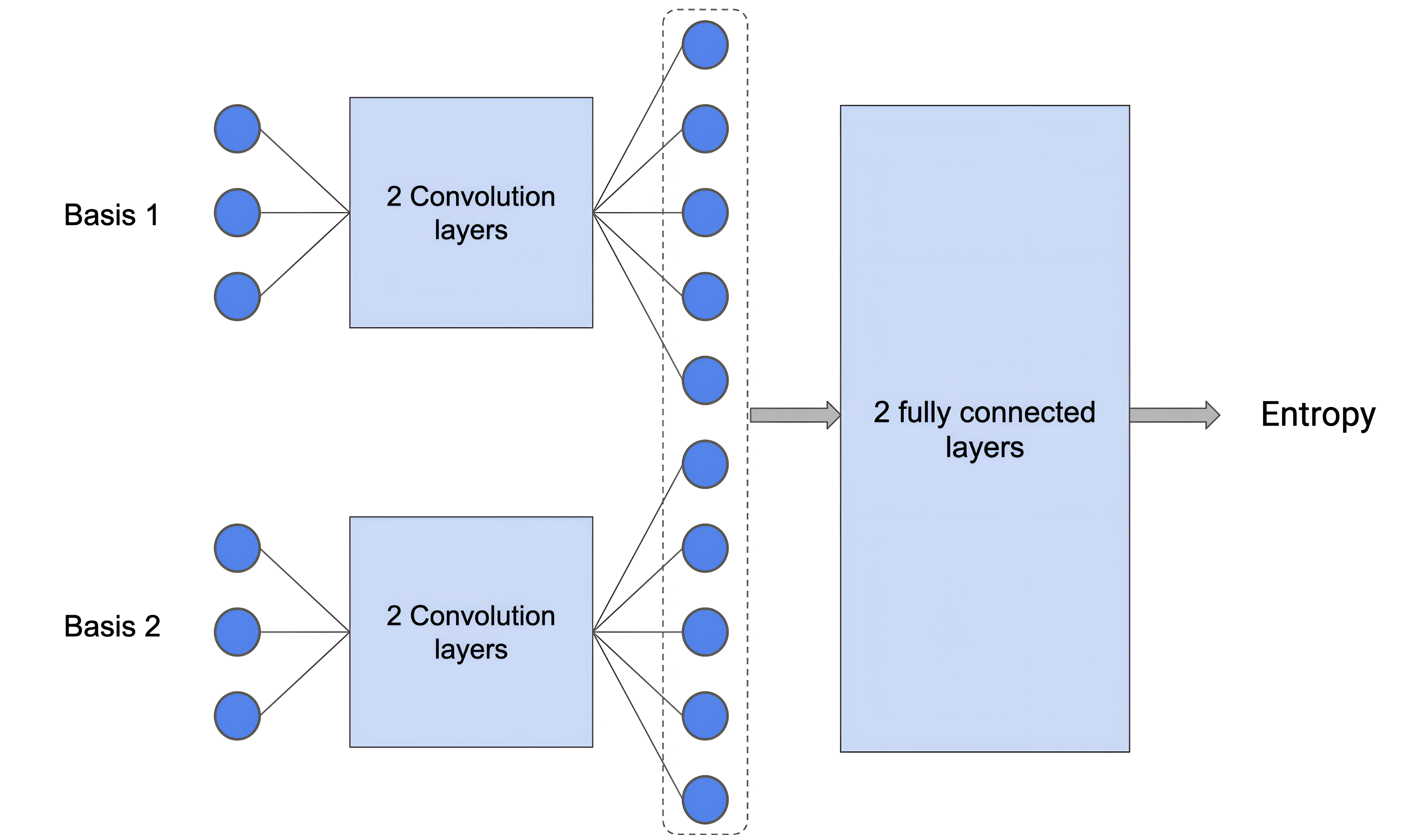}
    \caption{Schematic of the CNN-based entropy estimator. Measurement outcome distributions from different bases are processed by convolutional layers and combined via fully connected layers to predict the entropy. Only two bases are shown for illustration.}
    \label{fig:cnn_schematic}
\end{figure*}

To isolate the role of entanglement, we conducted an ablation study using A1 with fixed parameter count, varying the number of entangling gates per layer from 2 to 6. For higher entanglement settings (4 and 6 gates), both distributed and concentrated configurations were considered, corresponding to entangling operations spread across layers or applied near the end of the circuit. This entanglement experiment was performed on the same qutrit state to ensure consistent evaluation conditions.

\subsubsection{Ansatz Design for Two Qutrits}

Based on the three-qutrit study, A9 and A10 were extended to the two-qutrit setting due to their reduced entangling-gate requirements without significant loss in accuracy. The local variational blocks were retained, while the entanglement structure was adapted to the smaller system.

In even-numbered layers, two entangling gates are applied to achieve full connectivity, while odd-numbered layers use a single entangling gate with alternating direction to maintain balanced information flow. As in the three-qutrit case, local GMR rotations are interleaved with entangling operations, and each layer begins and ends with fixed basis-change gates.

\subsection{CNN-Based Entropy Estimation}

As a complementary alternative to VQAs, we consider a fully classical approach for estimating the von Neumann entropy directly from measurement data. Instead of reconstructing the density matrix or its spectrum, the model learns a mapping from measurement outcome probabilities to entropy values. This enables entropy estimation without explicit eigenvalue reconstruction, as spectral information is implicitly encoded in measurement statistics across bases.

Training and evaluation are performed using synthetically generated mixed multi-qutrit states, with exact entropies computed from the density matrix eigenvalues as ground-truth labels. The model is trained on ideal measurement probabilities without finite-shot noise.

\subsubsection{Measurement Representation}

The input to the model consists of measurement outcome probability distributions obtained from multiple bases, enabling entropy estimation directly from experimentally accessible data.

For entropy estimation in multi-qutrit systems, measurements are performed using tensor products of single-qutrit MUBs constructed from Gell--Mann operators. For a single qutrit ($d=3$), a complete set of four MUBs $\{\mathcal{B}_0,\dots,\mathcal{B}_3\}$ satisfies
\[
|\langle \psi | \phi \rangle|^2 = \frac{1}{3}, \quad \forall\, |\psi\rangle \in \mathcal{B}_i,\ |\phi\rangle \in \mathcal{B}_j,\ i \neq j,
\]
ensuring uniform overlap between bases and sensitivity to both population and coherence information.

Extending such constructions to global MUBs for composite qutrit systems is nontrivial, and practical implementations remain limited~\cite{lawrence2004qutrit}. We therefore employ tensor-product MUBs~\cite{mcnulty2016productmub}, defined for an $N$-qutrit system as
\[
\mathcal{B}_{\mu} = \mathcal{B}_{\mu_1} \otimes \cdots \otimes \mathcal{B}_{\mu_N}, \quad \mu_k \in \{0,1,2,3\}.
\]

Although these bases are not globally mutually unbiased, they provide a scalable and experimentally feasible measurement strategy that retains sensitivity to both local and multi-qutrit correlations relevant for entropy estimation.

In practice, a reduced subset of these bases is used to limit measurement cost while retaining sufficient information for entropy estimation.

Measurement distributions from each basis are processed independently by convolutional layers, and the resulting features are combined using fully connected layers to produce the entropy estimate (Fig.~\ref{fig:cnn_schematic}).

\subsubsection{Data Generation}

Training data consists of randomly generated mixed quantum states obtained by sampling eigenvalues from a Dirichlet distribution and applying Haar-random unitary rotations~\cite{zyczkowski2001induced}. To ensure uniform coverage across the entropy range, an entropy-stratified sampling strategy is used.

For each state, measurement probabilities are computed using a fixed subset of tensor-product MUBs. The same measurement configuration is used across all samples for a given system size, enabling consistent training and evaluation.

\subsubsection{Training Protocol}

Separate CNN models are trained for each system size. While the overall architectural design is kept consistent, model capacity is increased with system size by scaling both the convolutional feature extractor and the fully connected layers. This accommodates the growth in Hilbert space dimension and the increased complexity of correlations across measurement bases.

The network outputs a single scalar entropy estimate. A sigmoid activation is used at the output layer to enforce bounded predictions, and targets are normalized by $\log(D)$, with $D = 3^N$, so that the regression task is defined over a fixed range $[0,1]$ across all system sizes.

Training is performed using the Adam optimizer with mean absolute error (L1 loss) between predicted and true normalized entropy values. L1 loss is preferred over mean squared error, as it maintains sensitivity to moderate deviations and leads to more stable training across different degrees of mixedness.

All models are trained on noiseless measurement probabilities computed directly from the density matrices, allowing the intrinsic performance of the learning approach to be evaluated independently of sampling noise.
\begin{figure}[t]
    \centering
    \includegraphics[width=\linewidth]{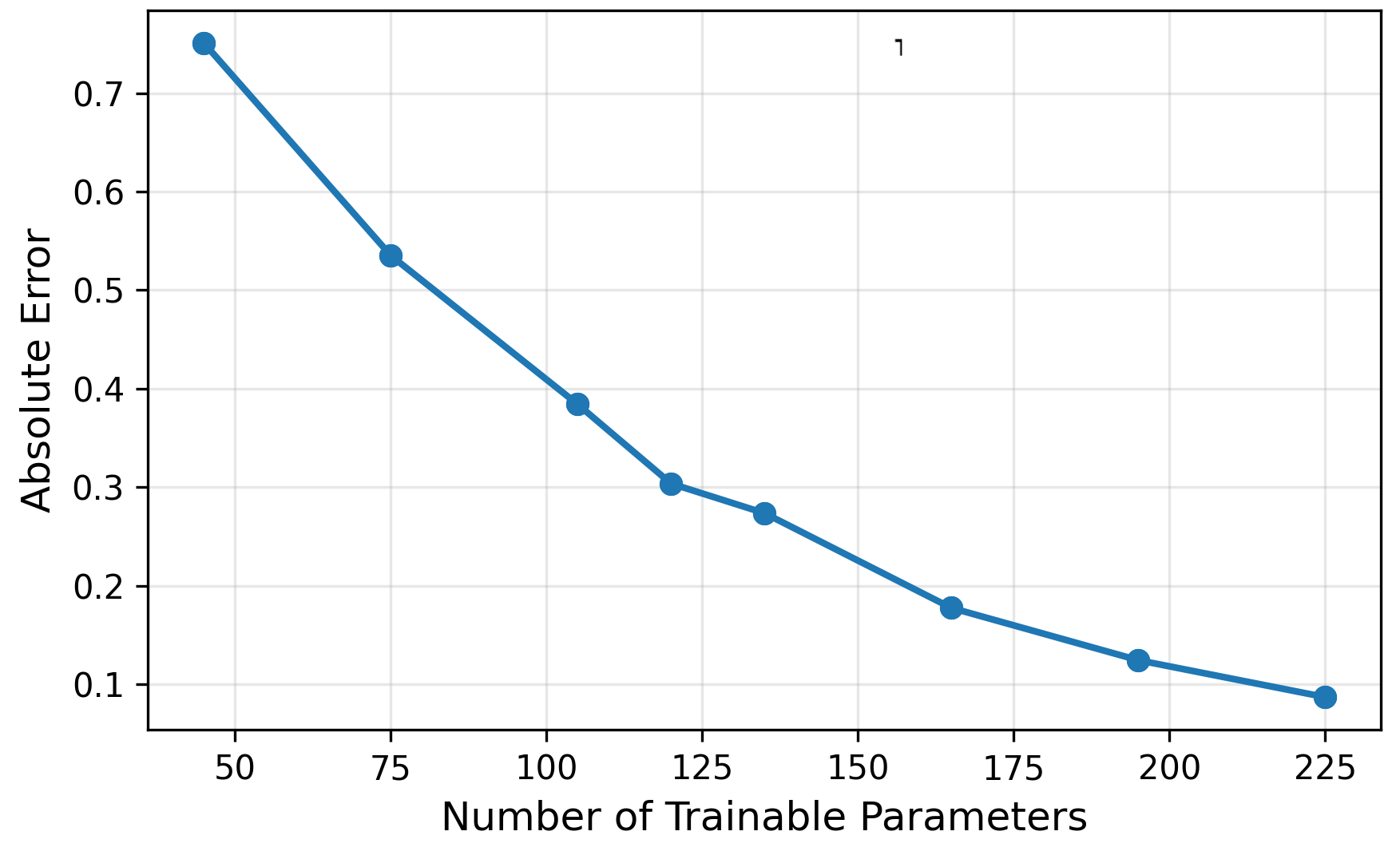}
    \caption{Parameter study using Ansatz~1 on a 60\% mixed 3-qutrit state. The entropy estimation error decreases with increasing number of trainable parameters under ideal simulation conditions.}
    \label{fig:parameter_study}
\end{figure}

\section{Results}
\subsection{VQA Results}
We quantify mixedness using the normalized entropy $\tilde{S}(\rho) = S(\rho)/\log D$. It ranges from $0$ for pure states to $1$ for maximally mixed states.

The analysis focuses on the three-qutrit system as the primary testbed.

The parameter sufficiency study uses Ansatz 1 on a 60\% mixed state. The number of trainable parameters is varied by increasing circuit layers. Each layer contributes 15 parameters. As shown in Fig.~\ref{fig:parameter_study}, the estimation error decreases with parameter count. Low-depth circuits show high error. This indicates that performance is mainly governed by the total number of parameters, provided sufficient entanglement is present. The trend holds under ideal simulation. On hardware, deeper circuits accumulate noise. This creates a trade-off between expressibility and depth.

\begin{figure}[t]
    \centering
    \includegraphics[width=\linewidth]{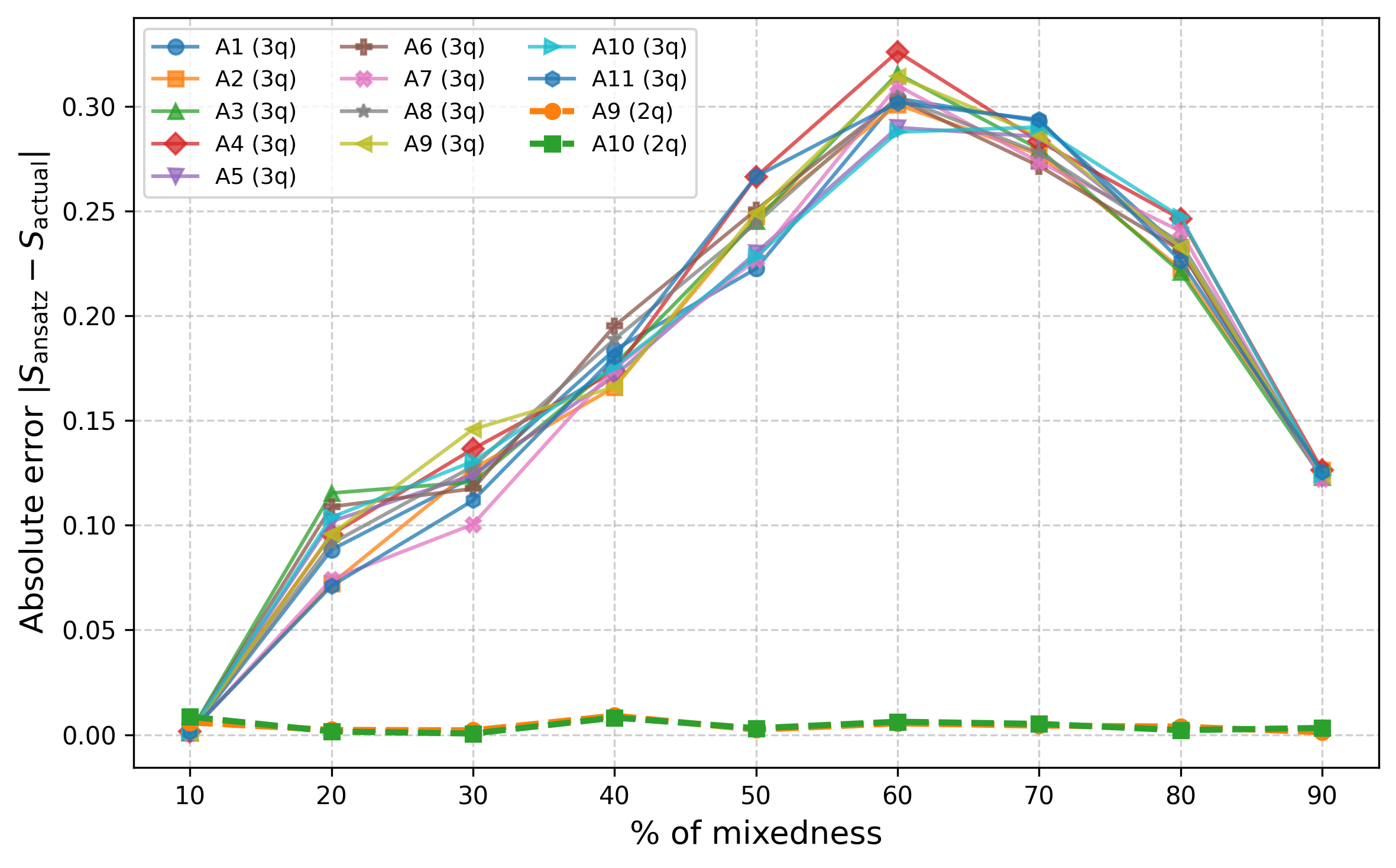}
    \caption{Absolute prediction error in the estimation of von Neumann entropy for two- and three-qutrit quantum states as a function of the percentage of mixedness. Two-qutrit states have negligible error while all ansatzes for three-qutrit VQA perform similar.}
    \label{fig:vqa_3qutrit_vn_error}
\end{figure}

Based on this observation, the parameter count is fixed for the three-qutrit experiments. We use approximately 120 parameters to isolate the effect of circuit architecture.

Figure~\ref{fig:vqa_3qutrit_vn_error} shows the entropy estimation error for three-qutrit states across different mixedness levels. For comparison, results for the two-qutrit system are also included in the same figure. The two-qutrit case achieves high accuracy across all mixedness levels, with worst-case error below $9\times10^{-3}$ nats. In contrast, the three-qutrit system shows higher error and a clear dependence on mixedness. The error peaks near $60\%$ mixedness at about $0.3$ nats. It decreases toward both low- and high-mixedness regimes. This reflects the increased difficulty of estimating entropy when the eigenvalue spectrum is neither sparse nor uniform.

Within the three-qutrit system, all ansatzes behave similarly. No architecture shows a consistent advantage. This suggests that accuracy is largely insensitive to circuit structure under a fixed parameter budget.

\begin{figure}[t]
    \centering
    \includegraphics[width=\linewidth]{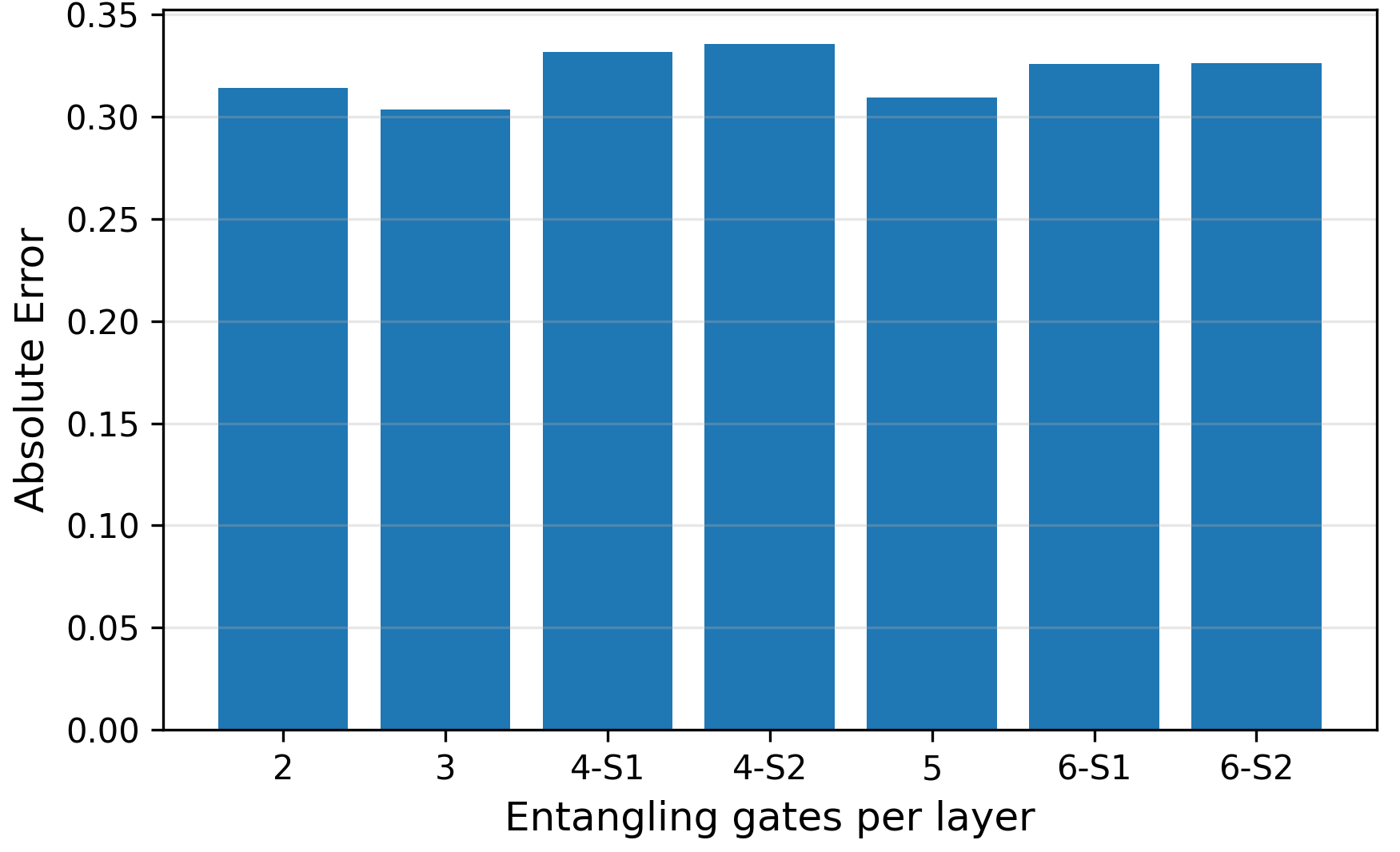}
    \caption{Entropy estimation error as a function of the number of entangling gates in three-qutrit case for A1 ansatz. Increasing entanglement beyond threshold has negligible effect.}
    \label{fig:entanglement_study}
\end{figure}

To study the role of entanglement, we vary the number of CSUM gates while keeping parameters fixed for the three-qutrit case. As shown in Fig.~\ref{fig:entanglement_study}, increasing entangling gates does not improve accuracy. The error saturates beyond a minimal threshold. Differences across configurations are negligible. This indicates that parameter count dominates once sufficient correlations are present. Additional entangling gates only increase circuit complexity.

From a practical perspective, architectures differ in efficiency. Ansatzes such as A9 and A10 achieve similar accuracy with fewer entangling gates. They are more suitable for implementation. Deeper designs, such as A11, increase circuit depth without improving performance.

To assess scalability, we extend the method to four-qutrit systems. We increase both depth and parameter count. The errors remain high, around $0.6$--$0.8$ nats. This indicates insufficient expressibility at practical depths. Further increases in depth are costly and noise-prone. These results highlight a key limitation of VQAs. They perform well for small systems but do not scale efficiently to larger qutrit systems.

Motivated by this limitation, we turn to classical learning-based approaches for higher-dimensional systems.

\subsection{CNN Results}

The CNN-based estimator is evaluated on two- to five-qutrit systems using synthetically generated mixed states with exact entropy labels.
\begin{figure}[t]
    \centering
    \includegraphics[width=\linewidth]{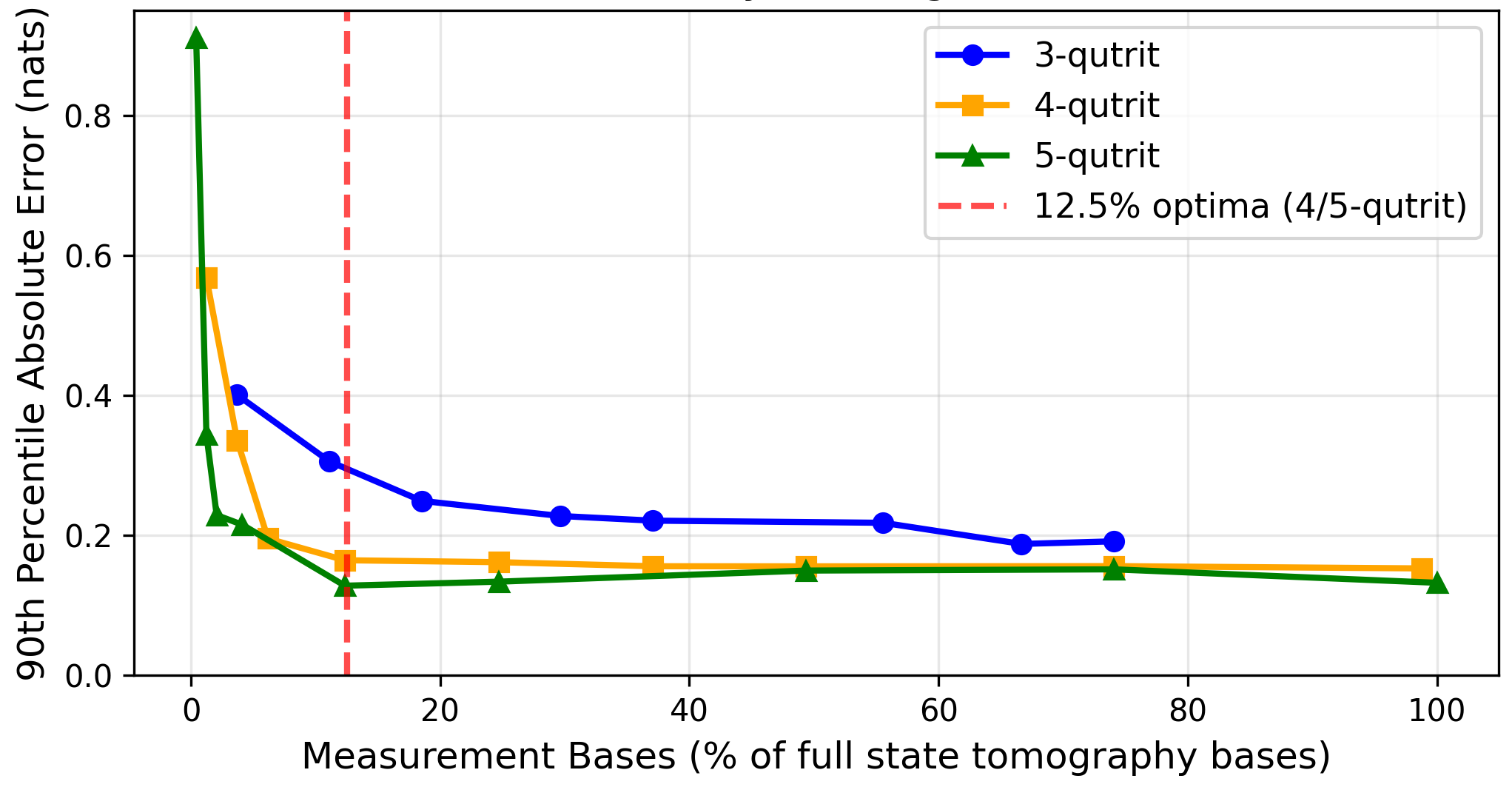}
    \caption{90th percentile error versus number of measurement bases for three-, four-, and five-qutrit systems. Approximately 12.5\% percent of full state tomography bases are sufficient to reach reasonable error for four- and five-qutrit systems}
    \label{fig:cnn_bases_study_all}
\end{figure}
Figure~\ref{fig:cnn_bases_study_all} shows the dependence of estimation error on the number of measurement bases. For four- and five-qutrit systems, accurate estimation is achieved with a small subset of bases: 10 bases for four qutrits (error approximately 0.16) and 30 bases for five qutrits (error approximately 0.13), beyond which performance saturates. For four- and five-qutrit systems, the full tomographic sets contain $81$ and $243$ bases, respectively. Thus, using 10 and 30 bases corresponds to $10/81 \approx 12.5\%$ and $30/243 \approx 12.5\%$, demonstrating strong measurement efficiency within the simulated setting considered.

For three-qutrit systems, performance improves steadily with additional bases, with diminishing returns beyond 8 bases and best accuracy achieved at 18 bases.

Based on these observations, a fixed subset of measurement bases is selected for each system size and used in all subsequent experiments.

\begin{figure}[t]
    \centering
    \includegraphics[width=\linewidth]{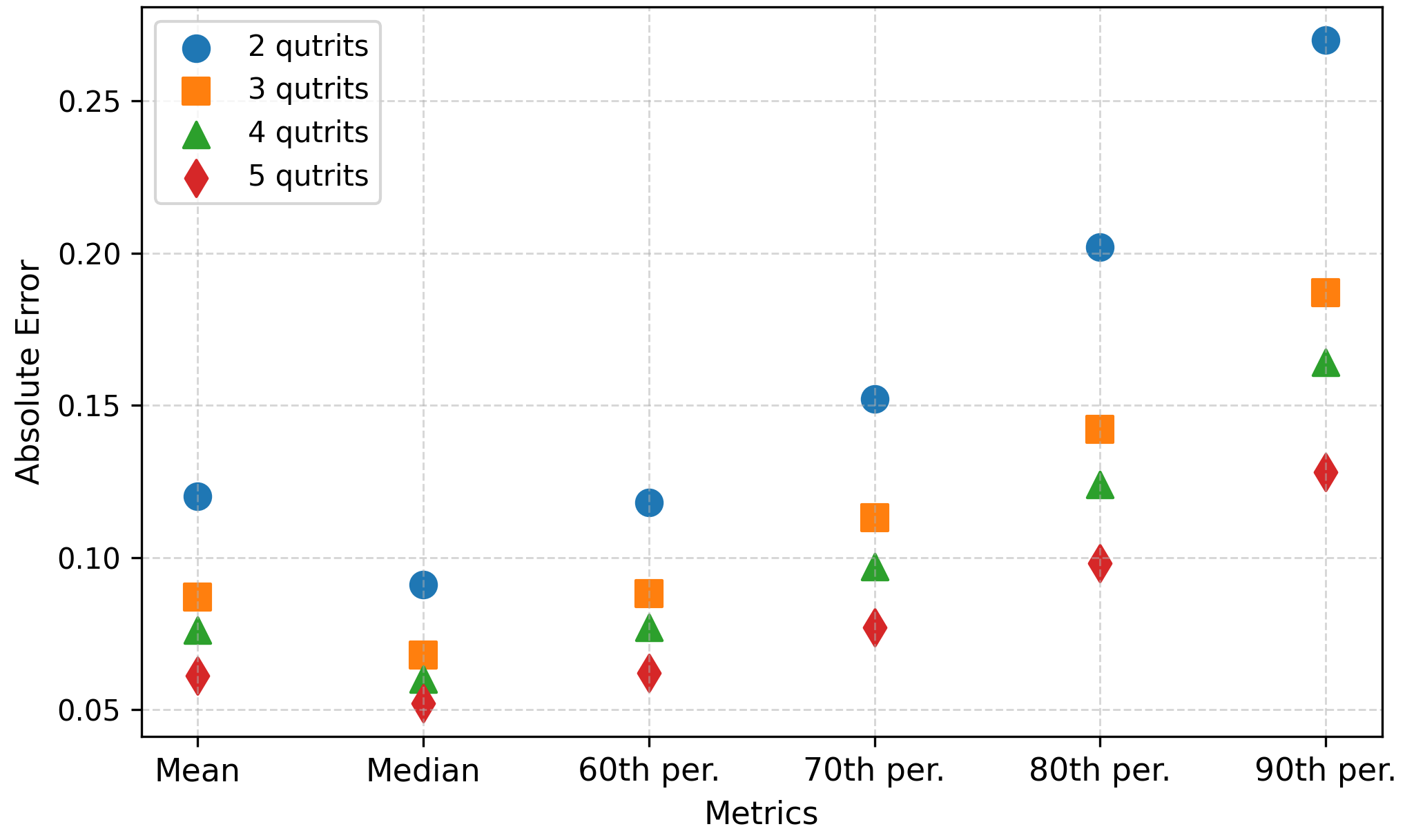}
    \caption{Prediction error statistics (mean, median, and percentiles) for von Neumann entropy estimation across system sizes.}
    \label{fig:cnn_vn_error_all_qutrits}
\end{figure}

Figure~\ref{fig:cnn_vn_error_all_qutrits} summarizes the prediction error statistics for von Neumann entropy estimation across 2–5 qutrit systems.

A clear improvement in accuracy is observed with increasing system size. The two-qutrit system exhibits the largest errors, with a 90th-percentile value of around $0.27$ nats. This reduces to approximately $0.19$ nats for three qutrits, around $0.16$ nats for four qutrits, and around $0.13$ nats for five qutrits.

This monotonic reduction in both central and tail errors indicates improved generalization in higher-dimensional systems. The larger errors in the two-qutrit case reflect the limited information content available from measurement statistics, whereas higher-dimensional systems provide richer structure that the model can exploit.

Overall, these results demonstrate strong scalability of the CNN-based estimator, with accuracy improving as system size increases.

\subsubsection{Out-of-Distribution Generalization}

\begin{figure}[t]
    \centering
    \includegraphics[width=\linewidth]{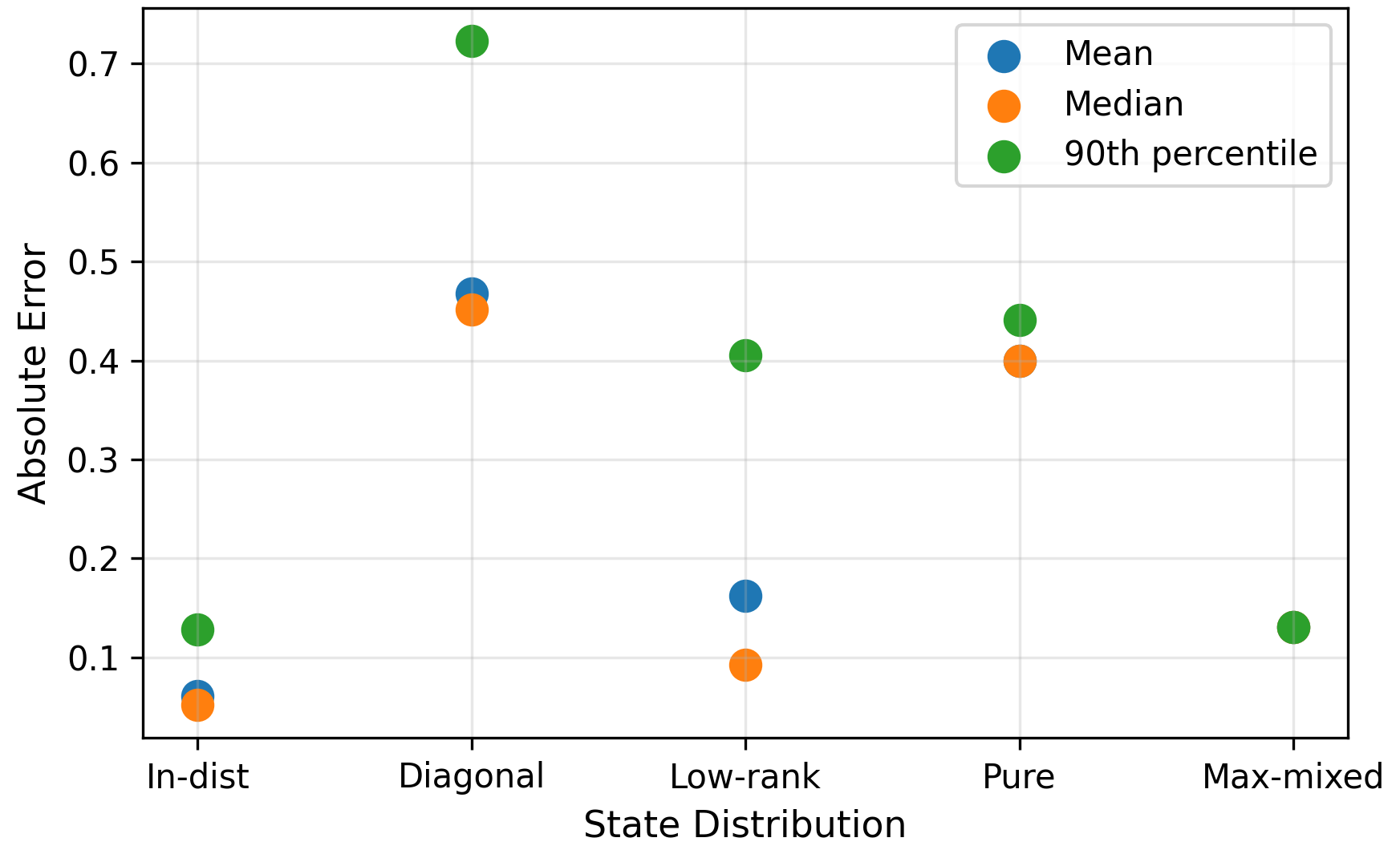}
    \caption{Out-of-distribution (OOD) performance for 5-qutrit von Neumann entropy estimation. The method generalizes well for OOD states too.}
    \label{fig:cnn_ood_generalization_5qutrit}
\end{figure}

Figure~\ref{fig:cnn_ood_generalization_5qutrit} evaluates out-of-distribution performance on structurally distinct quantum states. For in-distribution data, the model achieves a mean error of approximately $ 0.06$ and a 90th-percentile error of approximately $0.13$.

Across OOD classes, performance varies with state structure: diagonal states exhibit the largest errors (mean around $0.47$, 90th percentile around $0.72$), while low-rank, pure, and maximally mixed states show substantially lower errors. Despite this variation, errors remain bounded, indicating reasonable generalization beyond the training distribution.

\subsubsection{Effect of Measurement Noise}

\begin{figure}[t]
    \centering
    \includegraphics[width=\linewidth]{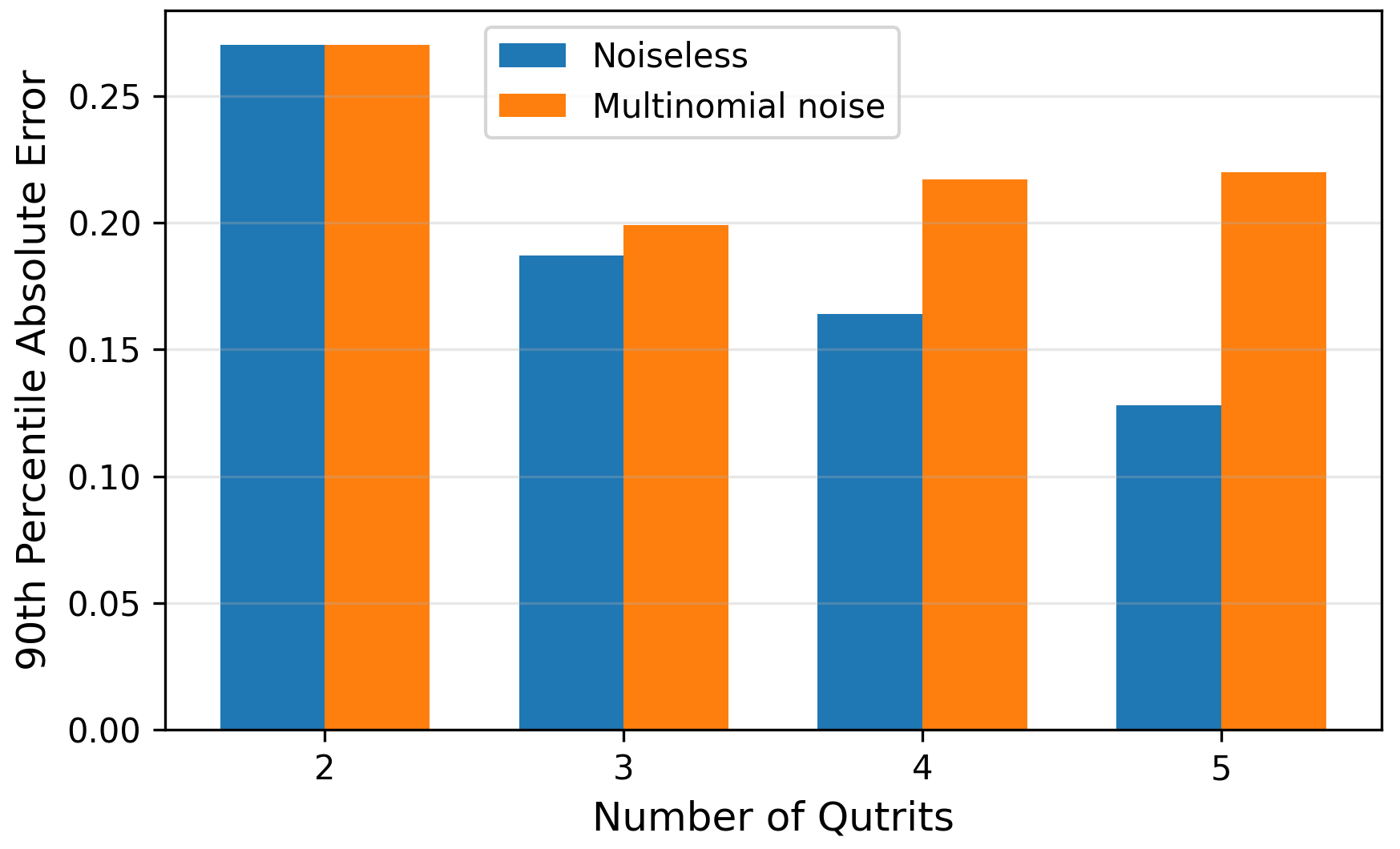}
    \caption{Effect of sampling noise on von Neumann entropy estimation.}
    \label{fig:cnn_noise}
\end{figure}

The impact of finite-shot measurement noise is evaluated by comparing performance under sampled measurement statistics as seen in Figure \ref{fig:cnn_noise}. For two-qutrit systems, the error remains largely unchanged, indicating low sensitivity to sampling noise. For larger systems, a modest increase in error is observed, but overall degradation remains limited, demonstrating robustness to measurement noise.

\section{Discussion}

A direct comparison between the VQA and CNN approaches can be made in the two- and three-qutrit regimes, where both methods are applicable. These regimes reveal a clear shift in relative performance with increasing system size.

For two-qutrit systems, the VQA significantly outperforms the CNN, achieving errors on the order of $10^{-3}$. This high accuracy arises from the ability of the variational circuit to approximately align the measurement basis with the eigenbasis of the density matrix, enabling near-optimal entropy estimation.

In contrast, the CNN exhibits larger errors in this low-dimensional setting, with a 90th-percentile error of around $ 0.27$. Without an explicit mechanism for eigenbasis alignment, the model must infer entropy directly from measurement statistics, which is more challenging when the underlying structure is simple.

In the three-qutrit case, this gap narrows substantially. The CNN achieves a 90th-percentile error of approximately $0.19$, while the VQA exhibits errors up to $0.3$ for comparable parameter budgets. Although VQA performance can be improved with additional parameters, this requires deeper circuits and more challenging optimization.

These results indicate a crossover in relative advantage. VQAs perform exceptionally well in low-dimensional systems due to their strong physical inductive bias, but their scalability is limited by circuit depth and optimization complexity. In contrast, the CNN improves with system size by learning entropy-relevant features directly from measurement data, without requiring explicit basis reconstruction.

Notably, in higher-dimensional systems, accurate CNN-based estimation is achieved using only a small fraction of measurement settings (as low as approximately $ 12.5\%$ of tomographic bases), highlighting its practical efficiency.

Overall, the two approaches exhibit complementary strengths: VQAs provide high accuracy and interpretability in small systems, while CNN-based methods offer superior scalability and robustness in larger systems. This suggests that system-size-dependent or hybrid strategies may provide an effective route for practical entropy estimation in multi-qutrit systems.

\section{Conclusion}

We studied entropy estimation in multi-qutrit systems using both variational quantum algorithms (VQAs) and classical convolutional neural networks (CNNs). For VQAs, a systematic analysis of ansatz families shows that estimation accuracy is primarily governed by the number of trainable parameters, while additional entanglement beyond a minimal threshold yields diminishing returns. Although highly accurate in low-dimensional systems, VQAs require rapidly increasing circuit depth and optimization effort as system size grows, limiting their scalability.

In contrast, the CNN-based approach enables direct entropy estimation from measurement statistics without explicit state reconstruction. Accurate predictions are achieved across larger systems using only a small fraction of measurement settings, demonstrating strong measurement efficiency and scalability within the simulated, noise-free setting considered.

Taken together, these results reveal a clear transition in performance: VQAs are well suited for small systems due to their physical interpretability, while classical learning-based methods become increasingly advantageous in higher-dimensional regimes. This highlights a fundamental trade-off between physical inductive bias and scalability in entropy estimation.

Future work will focus on incorporating realistic noise, optimizing measurement strategies, extension to r\'enyi entropy and to larger qudit systems. These directions will be essential for translating the proposed approaches to practical quantum platforms.

\section{Data availability}
The data that support the findings of this study are available from the authors upon request.

\section{Funding}
S.S.G. and A.P. acknowledge support from the Mphasis F1 Foundation, an Institute of Excellence grant funded by the Ministry of Education. The Qudit library~\cite{seksaria2025qudit} was also developed with this support.

\bibliographystyle{IEEEtran}
\bibliography{bibilography_uniform}

\end{document}